\documentclass[aps,prl,twocolumn,amsmath,amssymb,superscriptaddress,showpacs]{revtex4}
\usepackage{graphicx}% Include figure files
\usepackage{dcolumn}% Align table columns on decimal point
\usepackage{bm}% bold math
\usepackage[abs]{overpic} % Letters on the figure
\usepackage{feynmf}%Feynman diagrams
\usepackage{color}

\preprint{draftv6.0} %for submission
   \newcommand{\ppbar}{\ensuremath{\bar{p}p}}                                                                                             %
   \newcommand{\Bd}{\ensuremath{B^{0}}}                                                                                                   %
   \newcommand{\Bu}{\ensuremath{B^{+}}}                                                                                                   %
   \newcommand{\Bs}{\ensuremath{B_{s}^{0}}}                                                                                               %
   \newcommand{\Bdpipi}{\ensuremath{\Bd\rightarrow\pi^{+}\pi^{-}}}                                                                       %
   \newcommand{\BdKpi}{\ensuremath{\Bd\rightarrow K^{+}\pi^{-}}}                                                                        %
   \newcommand{\BsKK}{\ensuremath{\Bs\rightarrow K^{+}K^{-}}}                                                                           %
   \newcommand{\BsKpi}{\ensuremath{\Bs\rightarrow K^{-}\pi^{+}}}                                                                        %
                          
   \newcommand{\Lbppi}{\ensuremath{\Lb \rightarrow p\pi^{-}}}   
   \newcommand{\LbpK}{\ensuremath{\Lb \rightarrow p K^{-}}}

   \newcommand{\DKpi}{\ensuremath{D^0\rightarrow K^-\pi^+}}      
   \newcommand{\aDKpi}{\ensuremath{\overline{D}^0\rightarrow K^+\pi^-}}             
   \newcommand{\Dpipi}{\ensuremath{D^0\rightarrow \pi^+\pi^-}}             
  \newcommand{\Lb}{\ensuremath{\Lambda^0_b}}

   \newcommand{\dedx}{\ensuremath{\mathit{dE/dx}}}                                                                              %
   \newcommand{\micron}{\ensuremath{\mu\mathrm{m}}}
   \newcommand{\pgev}{\ensuremath{\mathrm{GeV}/c}}
   \newcommand{\ptot}{\ensuremath{p_{\mathrm{tot}}}}   
   \newcommand{\massgev}{\ensuremath{\mathrm{GeV}/c^{2}}}

\newcommand{\babar}{\mbox{\sl B\hspace{-0.4em} {\scriptsize\sl A}\hspace{-0.37em} \sl B\hspace{-0.4em} {\scriptsize\sl A\hspace{-0.02em}R}}}

\newcommand{\lumifb}{ fb$^{-1}$}
\newcommand{\like}{\ensuremath{\mathcal{L}}}

\newcommand{\mpipi}{\ensuremath{m_{\pi^+\pi^-}}}

\newcommand{\stat}{\ensuremath{\mathrm{~(stat)}}}		%	(stat.)
\newcommand{\syst}{\ensuremath{\mathrm{~(syst)}}}		%	(syst.)

\newcommand{\Lxy}{\ensuremath{L_{T}}}			%	Lxy
\newcommand{\CP}{\ensuremath{\mathit{CP}}}

\newcommand{\acp}{\ensuremath{\mathcal{A}}}
\newcommand{\acpraw}{\ensuremath{\tilde{\mathcal{A}}}}

\newcommand{\btof}{\ensuremath{b \to f}}
\newcommand{\abtoaf}{\ensuremath{\bar{b} \to \bar{f}}}
\newcommand{\acpbtof}{\ensuremath{\acp (b \to f)}}
\newcommand{\ACPuncorr}{\ensuremath{\acpraw = [N_{\btof} -  N_{\abtoaf}]/[N_{\btof} +  N_{\abtoaf}]}}

\newcommand{\ACPRAWdef}{\ensuremath{{\frac{\Gamma (\btof)-\Gamma(\abtoaf)}{\Gamma (\btof)+\Gamma (\abtoaf)}}}}

\begin{document}

%-----------------------------------------
%                    
%   Title of paper   
%                    
%-----------------------------------------

\title{Measurements of Direct \CP-Violating Asymmetries in Charmless Decays of Bottom Baryons}

%\input{September2011_Authors_a.tex}
%\input{September2011_Authors_new.tex}
%--------------------------------------------------------------
%\input{November2012_Authors.tex}
% Last update: $Date: 2014/02/17 18:16:27 $
\affiliation{Institute of Physics, Academia Sinica, Taipei, Taiwan 11529, Republic of China}
\affiliation{Argonne National Laboratory, Argonne, Illinois 60439, USA}
\affiliation{University of Athens, 157 71 Athens, Greece}
\affiliation{Institut de Fisica d'Altes Energies, ICREA, Universitat Autonoma de Barcelona, E-08193, Bellaterra (Barcelona), Spain}
\affiliation{Baylor University, Waco, Texas 76798, USA}
\affiliation{Istituto Nazionale di Fisica Nucleare Bologna, \ensuremath{^{ii}}University of Bologna, I-40127 Bologna, Italy}
\affiliation{University of California, Davis, Davis, California 95616, USA}
\affiliation{University of California, Los Angeles, Los Angeles, California 90024, USA}
\affiliation{Instituto de Fisica de Cantabria, CSIC-University of Cantabria, 39005 Santander, Spain}
\affiliation{Carnegie Mellon University, Pittsburgh, Pennsylvania 15213, USA}
\affiliation{Enrico Fermi Institute, University of Chicago, Chicago, Illinois 60637, USA}
\affiliation{Comenius University, 842 48 Bratislava, Slovakia; Institute of Experimental Physics, 040 01 Kosice, Slovakia}
\affiliation{Joint Institute for Nuclear Research, RU-141980 Dubna, Russia}
\affiliation{Duke University, Durham, North Carolina 27708, USA}
\affiliation{Fermi National Accelerator Laboratory, Batavia, Illinois 60510, USA}
\affiliation{University of Florida, Gainesville, Florida 32611, USA}
\affiliation{Laboratori Nazionali di Frascati, Istituto Nazionale di Fisica Nucleare, I-00044 Frascati, Italy}
\affiliation{University of Geneva, CH-1211 Geneva 4, Switzerland}
\affiliation{Glasgow University, Glasgow G12 8QQ, United Kingdom}
\affiliation{Harvard University, Cambridge, Massachusetts 02138, USA}
\affiliation{Division of High Energy Physics, Department of Physics, University of Helsinki, FIN-00014, Helsinki, Finland; Helsinki Institute of Physics, FIN-00014, Helsinki, Finland}
\affiliation{University of Illinois, Urbana, Illinois 61801, USA}
\affiliation{The Johns Hopkins University, Baltimore, Maryland 21218, USA}
\affiliation{Institut f\"{u}r Experimentelle Kernphysik, Karlsruhe Institute of Technology, D-76131 Karlsruhe, Germany}
\affiliation{Center for High Energy Physics: Kyungpook National University, Daegu 702-701, Korea; Seoul National University, Seoul 151-742, Korea; Sungkyunkwan University, Suwon 440-746, Korea; Korea Institute of Science and Technology Information, Daejeon 305-806, Korea; Chonnam National University, Gwangju 500-757, Korea; Chonbuk National University, Jeonju 561-756, Korea; Ewha Womans University, Seoul, 120-750, Korea}
\affiliation{Ernest Orlando Lawrence Berkeley National Laboratory, Berkeley, California 94720, USA}
\affiliation{University of Liverpool, Liverpool L69 7ZE, United Kingdom}
\affiliation{University College London, London WC1E 6BT, United Kingdom}
\affiliation{Centro de Investigaciones Energeticas Medioambientales y Tecnologicas, E-28040 Madrid, Spain}
\affiliation{Massachusetts Institute of Technology, Cambridge, Massachusetts 02139, USA}
\affiliation{University of Michigan, Ann Arbor, Michigan 48109, USA}
\affiliation{Michigan State University, East Lansing, Michigan 48824, USA}
\affiliation{Institution for Theoretical and Experimental Physics, ITEP, Moscow 117259, Russia}
\affiliation{University of New Mexico, Albuquerque, New Mexico 87131, USA}
\affiliation{The Ohio State University, Columbus, Ohio 43210, USA}
\affiliation{Okayama University, Okayama 700-8530, Japan}
\affiliation{Osaka City University, Osaka 558-8585, Japan}
\affiliation{University of Oxford, Oxford OX1 3RH, United Kingdom}
\affiliation{Istituto Nazionale di Fisica Nucleare, Sezione di Padova, \ensuremath{^{jj}}University of Padova, I-35131 Padova, Italy}
\affiliation{University of Pennsylvania, Philadelphia, Pennsylvania 19104, USA}
\affiliation{Istituto Nazionale di Fisica Nucleare Pisa, \ensuremath{^{kk}}University of Pisa, \ensuremath{^{ll}}University of Siena, \ensuremath{^{mm}}Scuola Normale Superiore, I-56127 Pisa, Italy, \ensuremath{^{nn}}INFN Pavia, I-27100 Pavia, Italy, \ensuremath{^{oo}}University of Pavia, I-27100 Pavia, Italy}
\affiliation{University of Pittsburgh, Pittsburgh, Pennsylvania 15260, USA}
\affiliation{Purdue University, West Lafayette, Indiana 47907, USA}
\affiliation{University of Rochester, Rochester, New York 14627, USA}
\affiliation{The Rockefeller University, New York, New York 10065, USA}
\affiliation{Istituto Nazionale di Fisica Nucleare, Sezione di Roma 1, \ensuremath{^{pp}}Sapienza Universit\`{a} di Roma, I-00185 Roma, Italy}
\affiliation{Mitchell Institute for Fundamental Physics and Astronomy, Texas A\&M University, College Station, Texas 77843, USA}
\affiliation{Istituto Nazionale di Fisica Nucleare Trieste, \ensuremath{^{qq}}Gruppo Collegato di Udine, \ensuremath{^{rr}}University of Udine, I-33100 Udine, Italy, \ensuremath{^{ss}}University of Trieste, I-34127 Trieste, Italy}
\affiliation{University of Tsukuba, Tsukuba, Ibaraki 305, Japan}
\affiliation{Tufts University, Medford, Massachusetts 02155, USA}
\affiliation{University of Virginia, Charlottesville, Virginia 22906, USA}
\affiliation{Waseda University, Tokyo 169, Japan}
\affiliation{Wayne State University, Detroit, Michigan 48201, USA}
\affiliation{University of Wisconsin, Madison, Wisconsin 53706, USA}
\affiliation{Yale University, New Haven, Connecticut 06520, USA}

\author{T.~Aaltonen}
\affiliation{Division of High Energy Physics, Department of Physics, University of Helsinki, FIN-00014, Helsinki, Finland; Helsinki Institute of Physics, FIN-00014, Helsinki, Finland}
\author{S.~Amerio\ensuremath{^{jj}}}
\affiliation{Istituto Nazionale di Fisica Nucleare, Sezione di Padova, \ensuremath{^{jj}}University of Padova, I-35131 Padova, Italy}
\author{D.~Amidei}
\affiliation{University of Michigan, Ann Arbor, Michigan 48109, USA}
\author{A.~Anastassov\ensuremath{^{v}}}
\affiliation{Fermi National Accelerator Laboratory, Batavia, Illinois 60510, USA}
\author{A.~Annovi}
\affiliation{Laboratori Nazionali di Frascati, Istituto Nazionale di Fisica Nucleare, I-00044 Frascati, Italy}
\author{J.~Antos}
\affiliation{Comenius University, 842 48 Bratislava, Slovakia; Institute of Experimental Physics, 040 01 Kosice, Slovakia}
\author{G.~Apollinari}
\affiliation{Fermi National Accelerator Laboratory, Batavia, Illinois 60510, USA}
\author{J.A.~Appel}
\affiliation{Fermi National Accelerator Laboratory, Batavia, Illinois 60510, USA}
\author{T.~Arisawa}
\affiliation{Waseda University, Tokyo 169, Japan}
\author{A.~Artikov}
\affiliation{Joint Institute for Nuclear Research, RU-141980 Dubna, Russia}
% ...

\collaboration{CDF Collaboration}
\altaffiliation[With visitors from]{
\ensuremath{^{a}}University of British Columbia, Vancouver, BC V6T 1Z1, Canada,
\ensuremath{^{b}}Istituto Nazionale di Fisica Nucleare, Sezione di Cagliari, 09042 Monserrato (Cagliari), Italy,
\ensuremath{^{c}}University of California Irvine, Irvine, CA 92697, USA,
\ensuremath{^{d}}Institute of Physics, Academy of Sciences of the Czech Republic, 182~21, Czech Republic,
\ensuremath{^{e}}CERN, CH-1211 Geneva, Switzerland,
\ensuremath{^{f}}Cornell University, Ithaca, NY 14853, USA,
\ensuremath{^{g}}University of Cyprus, Nicosia CY-1678, Cyprus,
\ensuremath{^{h}}Office of Science, U.S. Department of Energy, Washington, DC 20585, USA,
\ensuremath{^{i}}University College Dublin, Dublin 4, Ireland,
\ensuremath{^{j}}ETH, 8092 Z\"{u}rich, Switzerland,
\ensuremath{^{k}}University of Fukui, Fukui City, Fukui Prefecture, Japan 910-0017,
\ensuremath{^{l}}Universidad Iberoamericana, Lomas de Santa Fe, M\'{e}xico, C.P. 01219, Distrito Federal,
\ensuremath{^{m}}University of Iowa, Iowa City, IA 52242, USA,
\ensuremath{^{n}}Kinki University, Higashi-Osaka City, Japan 577-8502,
\ensuremath{^{o}}Kansas State University, Manhattan, KS 66506, USA,
\ensuremath{^{p}}Brookhaven National Laboratory, Upton, NY 11973, USA,
\ensuremath{^{q}}Queen Mary, University of London, London, E1 4NS, United Kingdom,
\ensuremath{^{r}}University of Melbourne, Victoria 3010, Australia,
\ensuremath{^{s}}Muons, Inc., Batavia, IL 60510, USA,
\ensuremath{^{t}}Nagasaki Institute of Applied Science, Nagasaki 851-0193, Japan,
\ensuremath{^{u}}National Research Nuclear University, Moscow 115409, Russia,
\ensuremath{^{v}}Northwestern University, Evanston, IL 60208, USA,
\ensuremath{^{w}}University of Notre Dame, Notre Dame, IN 46556, USA,
\ensuremath{^{x}}Universidad de Oviedo, E-33007 Oviedo, Spain,
\ensuremath{^{y}}CNRS-IN2P3, Paris, F-75205 France,
\ensuremath{^{z}}Universidad Tecnica Federico Santa Maria, 110v Valparaiso, Chile,
\ensuremath{^{aa}}The University of Jordan, Amman 11942, Jordan,
\ensuremath{^{bb}}Universite catholique de Louvain, 1348 Louvain-La-Neuve, Belgium,
\ensuremath{^{cc}}University of Z\"{u}rich, 8006 Z\"{u}rich, Switzerland,
\ensuremath{^{dd}}Massachusetts General Hospital, Boston, MA 02114 USA,
\ensuremath{^{ee}}Harvard Medical School, Boston, MA 02114 USA,
\ensuremath{^{ff}}Hampton University, Hampton, VA 23668, USA,
\ensuremath{^{gg}}Los Alamos National Laboratory, Los Alamos, NM 87544, USA,
\ensuremath{^{hh}}Universit\`{a} degli Studi di Napoli Federico I, I-80138 Napoli, Italy
}
\noaffiliation
% Last update: $Date: 2014/02/17 18:16:27 $
%--------------------------------------------------------------

\date{December 9, 2014}
%\date{\today}
%%%%%%%%%%%%%%%%
%              %
%   Abstract   %
%              %
%%%%%%%%%%%%%%%%
\begin{abstract}
We report final measurements of direct \CP--violating asymmetries in
charmless decays of neutral bottom hadrons to pairs of charged hadrons
with the upgraded Collider Detector at the Fermilab Tevatron.  
Using the complete $\sqrt{s}=1.96$~TeV  %$p\bar{p}$ 
proton-antiproton collisions
data set, corresponding to \mbox{9.3\lumifb} of integrated luminosity,
we measure $\acp(\Lbppi) = +0.06 \pm 0.07\stat \pm 0.03\syst$
and $\acp(\LbpK) = -0.10 \pm 0.08\stat \pm 0.04\syst$, compatible with no %\CP\ 
asymmetry. 
In addition we measure the \CP--violating asymmetries in \BsKpi\ and \BdKpi\ decays to be
$\acp(\BsKpi) = +0.22 \pm 0.07\stat \pm 0.02\syst$ and $\acp(\BdKpi) = -0.083\pm 0.013 \stat \pm 0.004\syst$,
respectively, which are significantly different from zero and consistent with current world averages.
%%%%%%%
% We report final measurements of direct $\mathit{CP}$--violating asymmetries in
% charmless decays of neutral bottom hadrons to pairs of charged hadrons
% with the upgraded Collider Detector at the Fermilab Tevatron.  
% Using the complete $\sqrt{s}=1.96$~TeV  proton-antiproton collisions
% data set, corresponding to \mbox{9.3 fb$^{-1}$} of integrated luminosity,
% we measure $\mathcal{A}( \Lambda^0_b \rightarrow p\pi^{-}) = +0.06 \pm 0.07\mathrm{~(stat)} \pm 0.03\mathrm{~(syst)}$
% and $\mathcal{A}(\Lambda^0_b \rightarrow pK^{-}) = -0.10 \pm 0.08\mathrm{~(stat)}  \pm 0.04\mathrm{~(syst)}$, compatible with no asymmetry. 
% In addition we measure the $\mathit{CP}$--violating asymmetries in  $B^0_s \rightarrow K^{-}\pi^{+}$ and $B^0 \rightarrow K^{+}\pi^{-}$  decays to be
% $\mathcal{A}( B^0_s \rightarrow K^{-}\pi^{+} ) = +0.22 \pm 0.07\mathrm{~(stat)}  \pm 0.02\mathrm{~(syst)}$ and 
% $\mathcal{A}(B^0 \rightarrow K^{+}\pi^{-} ) = -0.083\pm 0.013 \mathrm{~(stat)} \pm 0.004\mathrm{~(syst)}$,
% respectively, which are significantly different from zero and consistent with current world averages.
\end{abstract}

%\keywords{}


\maketitle

%\section{}
%\section{Introduction\label{sec:Intro}}
%

%-------------------------------------------------------------------------------------------
%                                      General introduction     
%-------------------------------------------------------------------------------------------

The experimentally established noninvariance of fundamental interactions under 
the combined symmetry transformations of charge conjugation and parity inversion (\CP\ violation) 
is described within the standard model (SM) %\CP\ symmetry is broken 
through the Cabibbo-Kobayashi-Maskawa (CKM) mechanism~\cite{Cabibbo-KM} %. %~\cite{Cabibbo,Kobayashi:1973fv}.
by the presence of a single complex phase %that appears 
%in the unitary matrix $V_{\rm CKM}$, describing three-generation quark mixing, allows breaking the \CP\ symmetry.
in the unitary three-generation quark-mixing matrix.
%4three-generation quark mixing unitary matrix. % $V_{\rm CKM}$. %, describing three-generation quark mixing, allows breaking the \CP\ symmetry.
All direct measurements of elementary particle phenomena to date %strongly 
support the CKM phase being the dominant source of \CP\ violation observed in quark transitions.
%However, widely accepted theoretical arguments and cosmological observations~\cite{Gavela:1993ts} suggest that
However, widely accepted theoretical arguments and cosmological observations suggest that
%However, there are good theoretical reasons to believe that 
%the SM cannot be the ultimate answer, 
%and there must be some new interactions at very high energies, 
the SM might be a lower-energy approximation of more generally valid theories
which are likely to possess a different  \CP\ structure
and therefore should manifest themselves as deviations from the CKM scheme.
%In addition, cosmological observations of matter--antimatter asymmetry in the universe
%are difficult to explain without assuming much larger amounts of \CP\
%violation than can be accommodated by the SM~\cite{Cohen:1993nk}. %\cite{Sakharov:1967dj}.

The decays of $b$ hadrons are highly relevant
in this context, with nonleptonic final states being particularly interesting.
They are sensitive to possible new
%This is due to the presence of 
contributions from internal \textit{loop} amplitudes, 
which provide a sensitive probe into energies higher than those 
accessible by direct searches. %Due to 
%Owing to the presence of hadronic factors in the decay amplitudes, 
Hadronic factors in the decay amplitudes make 
accurate SM predictions for individual decays %are 
difficult to obtain. Hence, the most useful information 
is obtained by combining multiple measurements 
of processes related by dynamical symmetries, allowing the cancellation of the unknown model parameters. 
An observable well suited for such studies is the \textit{direct} \CP\  asymmetry~\cite{coniugate}
%is  It shows up, 
%can be experimentally observed, 
% for non-CP-symmetric final states, if the partial decay-width ($\Gamma$) 
%of a particle into a final state differs from the width 
%of the corresponding antiparticle into the \CP-conjugate final state. For a generic $b-$hadron decay $b \to f$, 
%with $f\neq\bar{f}$, direct \CP-violating asymmetry is defined
%as~\cite{coniugate}
\begin{equation}
\mathcal{A}=\ACPRAWdef. 
\end{equation}
where $\Gamma$  is the partial decay-width of a generic $b-$hadron decay ($b \to f$)
 with non-CP-symmetric final state $f\neq\bar{f}$.
%where $c_{f} = \varepsilon(f)/\varepsilon(\bar{f})$ is the ratio
%between the efficiencies for triggering and reconstructing 
%the final states $f$ and $\bar{f}$. 
Recent examples of interplay between different measurements include the significant difference observed between the measured direct 
 \CP\ asymmetries for \BdKpi\ and $\Bu\to K^+\pi^0$ decays~\cite{acp_bfactories}, which prompted intense experimental and theoretical 
searches for an explanation,
either by an enhanced color-suppressed SM \textit{tree} contribution~\cite{Gronau:2005kz}, or by %new physics in the
non-SM physics in the
electroweak penguin loop~\cite{Khalil:2009zf}. Similarly, the comparison of the direct \CP\ asymmetries in \BsKpi\ and \BdKpi\ decays has been 
investigated as a nearly model-independent test for the presence of non-SM physics~\cite{Gronau:2000md,Gronau:2013mda}, and has
been experimentally performed only very recently~\cite{Aaij:2013iua}.

%\textcolor{red}{For non-CP-symmetric final states violation of \CP\ is \textit{direct} 
%if the partial decay-width %($\Gamma$) 
%of a particle into a final state differs from the width 
%of the corresponding antiparticle into the \CP-conjugate final state.}

%------------------------------------
%   Specific Intro   
%------------------------------------

While the properties of %these and other 
$b$ mesons decays have been studied in detail and no deviation from
the SM has yet been conclusively established, the decays of $b$ baryons
are still largely unexplored. 
An accurate experimental investigation of their \CP\ asymmetries is useful %indispensable 
to complete the current picture of charmless decays of $b$ hadrons. 
%The \LbpK\ and \Lbppi\ decays proceed through the same weak
%transitions of their corresponding two-body charmless hadronic $b$--meson decays. 
The \LbpK\ and \Lbppi\ decays proceed through the same weak
transitions as the corresponding two-body charmless hadronic $b$--meson decays. 
The first measurements~\cite{Aaltonen:2008hg} of their branching
fractions were not well described by predictions~\cite{Mohanta:2000nk}. 
In particular, the measured ratio of 
branching fractions $\mathcal{B}(\Lbppi)/\mathcal{B}(\LbpK)=0.66 \pm 0.14\stat \pm 0.08\syst$
significantly deviated from the predicted 
value of $2.6^{+2.0}_{-0.5}$~\cite{Lu:2009cm}. The  discrepancy  %low value 
has been recently confirmed by an 
independent measurement from the LHCb Collaboration~\cite{Aaij:2012as}. %hep-ex:1206.2794
Since branching ratios are potentially sensitive to new physics contributions~\cite{Wang:2013roa,Mohanta:2010eb}, %hep-ex:1010.1152
further investigation is clearly important~\cite{Gronau:2013mza}. The same
 calculations of Ref.~\cite{Lu:2009cm} also predict \CP\ asymmetries up to 30\%, 
%that the precision of previous measurements did not allow testing. 
which were not testable by the previous measurements.

%In this Letter we report on measurements of direct~\cite{dcpv} \CP\ violation
In this Letter we report on measurements of direct \CP\ violation
in two-body charmless decays of bottom baryons and mesons performed using the full data set
collected by the upgraded Collider Detector (CDF II) at the Fermilab Tevatron, corresponding to 
9.3\lumifb\ of integrated luminosity from
\ppbar\ collisions at $\sqrt{s} = 1.96$ TeV. 
This is an update of a previous measurement based 
on a subsample of the present data~\cite{Aaltonen:2011qt} and provides % both
significantly improved measurements of the baryonic
decay modes \LbpK\ and \Lbppi\ which are unique. 
We also present final measurements on the meson decay modes \BsKpi\ and \BdKpi. 

%---------------------------------------
%   Detector   
%---------------------------------------

The CDF II detector is a multipurpose magnetic spectrometer surrounded by calorimeters and muon detectors. 
%It is described in detail in Ref.~\cite{Acosta:2004yw}, and the detector subsystems relevant for this 
%analysis are discussed in Ref.~\cite{Abulencia:2006psa}.
The detector subsystems relevant for this 
analysis are discussed in Ref.~\cite{Acosta:2004yw,Abulencia:2006psa}.
Data are collected by a three-level on-line event-selection system (trigger).
At level~1, charged-particle trajectories (tracks) are reconstructed 
in the plane transverse to the beam line~\cite{CDF-coordinates}.
Two oppositely-charged particles are required with reconstructed transverse momenta $p_{T1}, p_{T2} > 2~\pgev$, 
a scalar sum $p_{T1}+p_{T2} > 5.5~\pgev$, and an azimuthal opening angle $\Delta\phi < 135^{\circ}$. 
At level~2, tracks are combined with silicon-tracking-detector measurement hits,
%and their impact parameter $d$ (transverse distance of closest approach to the beam line) 
and the impact parameter $d$ (transverse distance of closest approach to the beam line) of each
is determined with 45 \micron\ resolution (including the beam spread) and is required to satisfy $0.1 < d < 1.0$ mm. 
A tighter opening-angle requirement, $20^{\circ} < \Delta\phi < 135^{\circ}$, is also applied.
Each track pair is then used to form a $b$--hadron candidate ($H_b=B^0,B^0_s,\Lb$) that is  
required to have an impact parameter $d_{H_b} < 140~\micron$ and to have traveled a 
distance $\Lxy > 200~\micron$ in the transverse plane. At level~3,  a cluster of computers confirms the selection with a full event reconstruction.

%---------------------------------------
%   Offline selection   
%---------------------------------------

\begin{figure}[tb]
\includegraphics[scale=0.35]{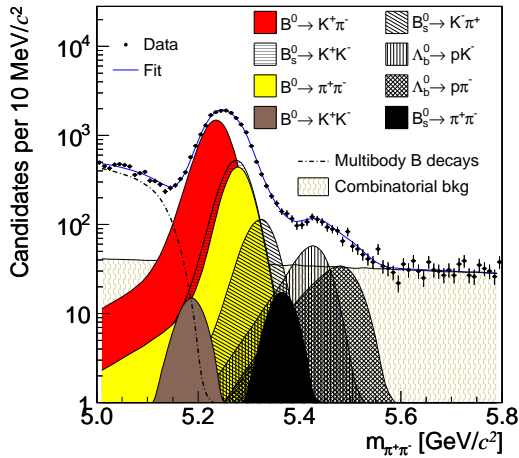}
\caption{\label{fig:projections} Mass distribution of reconstructed candidates, where the charged pion mass is assigned to both tracks.
The sum of the fitted distributions and the individual components (C-conjugate decay modes are also implied) of signal and background are overlaid on the data distribution.  }
\end{figure}

The offline selection is based on a more accurate 
determination of the same quantities used in the trigger with the 
addition of two further observables: 
the isolation of the $H_b$ candidate~\cite{Aaltonen:2008hg}  %~\cite{Isolation},
and the quality of the three-dimensional fit ($\chi^{2}$  with one degree of freedom) of the 
%decay vertex~\cite{vertex_def} of the $B$ candidate. 
candidate decay vertex.
%$B$ candidate decay vertex. 
We use the selection originally devised for the \BsKpi\ search~\cite{Aaltonen:2008hg}.
At most one $H_b$ candidate per event is found,  for which the invariant mass $m_{\pi^+\pi^-}$  is calculated using 
a charged-pion mass assignment for both decay products. 
The resulting mass distribution is shown in Fig.~\ref{fig:projections}. 
It is dominated by the overlapping contributions of the \BdKpi, \Bdpipi, and \BsKK\ 
decays~\cite{Abulencia:2006psa,Aaltonen:2011qt} with backgrounds 
from misreconstructed multi-body $b$--hadron decays (physics background) and random pairs of charged particles (combinatorial background).
Signals for the \BsKpi, \Lbppi, and \LbpK\ decays 
populate masses higher than the prominent narrow structure
(5.33--5.55 \massgev)~\cite{Aaltonen:2008hg}.
The final data sample consists of 28~230 $H_b$ candidates.

%---------------------------------------
%   Signal separation   
%---------------------------------------

We use an extended unbinned maximum likelihood fit incorporating kinematic (kin) 
and particle-identification (PID) information, to disentangle the various contributions. 
From the fit we determine the %fraction of each %individual mode
fraction of events from each decay mode 
and the asymmetries, uncorrected for
instrumental effects, \ACPuncorr, of the flavor-specific decays \BdKpi, \BsKpi, \Lbppi, and \LbpK. 
For each channel, $N_{\btof}(N_{\abtoaf})$ is the number of reconstructed decays of the hadron containing
the $b(\bar{b})$ quark into the final state $f(\bar{f})$, where the flavor of the hadron 
is inferred from the charges of the final-state particles.
In evaluating asymmetries we neglect any effect from 
\CP\ violation in $b$--meson flavor mixing~\cite{PDG}.  Production 
asymmetries also have negligible effects, as 
in $\bar{p}p$ collisions $b$ and $\bar{b}$ quarks are produced in equal numbers %,
%being the products of \CP-conserving strong  interactions. 
%In addition 
and the symmetry in pseudorapidity of the CDF~II detector, at level of 1\%. %(1.15 +/- 0.05)%.
This ensures equal acceptance down to a level of $10^{-3}$ even in the presence of possible %an unlikely large 10\% % possible 
forward-backward production asymmetries, constrained by \CP\ conservation to change sign 
for opposite values of pseudorapidity. Detailed studies performed on large samples of 
$D^0$ two-body  decays show residual effects on the \CP-asymmetry measurements  
of the order of  $10^{-4}$ ~\cite{Aaltonen:2011se} .
%being well below few per cent level~\cite{Grinstein:2013mia}. 
%\textcolor{red}{In addition the geometrical symmetries of the CDF~II detector 
%ensure equal acceptance even in the presence of a possible forward-backward production 
%asymmetry.}
%We also  assume that any effect from 
%\CP\ violation in $b$--meson flavor mixing is negligible~\cite{PDG}. 
The likelihood is defined as
$\mathcal{L} =  \frac{\nu^{N}}{N!}e^{-\nu} \prod^{N}_{i=1}  \mathcal{L}_i$
where $N$ is the total number of observed $H_b$ candidates, $\nu$ is the estimator of $N$ to be determined by the fit, 
 and  the likelihood for the $i$th event is
\begin{eqnarray}\label{eq:likelihood}
    \mathcal{L}_i & = & (1-b)\sum_{j} f_j \mathcal{L}^{\mathrm{kin}}_j  \mathcal{L}^{\mathrm{PID}}_j \nonumber \\
                  &   & +  b \left[ f_{\rm{p}} \mathcal{L}^{\mathrm{kin}}_{\mathrm{p}}
    \mathcal{L}^{\mathrm{PID}}_{\mathrm{p}}+
   (1-f_{\rm{p}}) \mathcal{L}^{\mathrm{kin}}_{\mathrm{c}}
    \mathcal{L}^{\mathrm{PID}}_{\mathrm{c}}
	\right], \label{like_fit}
%	\right),
\end{eqnarray}
where the index $j$ runs over all signal decay modes, 
 and the index `p' (`c') labels the physics (combinatorial) 
background term. The $f_j$ are signal fractions to be determined by 
the fit, together with the background fraction parameters $b$ and $f_{\rm{p}}$.
$\like^{\mathrm{kin}}_{j,\mathrm{p,c}}$  and  $\like^{\mathrm{PID}}_{j\mathrm{,p,c}}$ are respectively the likelihood terms incorporating
the kinematic and PID information for signal decay modes and backgrounds, defined in more detail later.
%-----------------------------------------------------------------------------------------
%                Inv mass figure
%-----------------------------------------------------------------------------------------
%-----------------------------------------------------------------------------------------

For each charged-hadron pair, the kinematic information is summarized by three loosely correlated
observables: the squared mass
$\mpipi^2$; the charged momentum asymmetry
$\beta = (p_{+}-p_{-})/(p_{+}+p_{-})$, where $p_+$ ($p_-$) is the magnitude of the momentum
of the positive (negative) particle; and  the scalar sum of particle momenta 
$\ptot=p_+ + p_-$.
These variables allow evaluation of the squared invariant mass of a candidate for  any
mass assignment of the positively- and negatively-charged decay products~\cite{Aaltonen:2011jv}.
% ($m_{a^+}$,$m_{b^-}$), using the equation
%\begin{eqnarray}\label{eq:Mpipi2}
% m^{2}_{a^+b^-}  =  m^{2}_{\pi^+\pi^-}  -  m_{\pi^+}^2  - m_{\pi^-}^2+ m_{a^+}^2+m_{b^-}^2 +
%                 \nonumber    \\
%          -  2 \sqrt{p_{+}^2+m_{\pi^+}^2} \sqrt{p_{-}^2+m_{\pi^-}^2}  
%                    +  2\sqrt{p_{+}^2+m_{a^+}^2} \sqrt{p_{-}^2+m_{b^-}^2},
% \end{eqnarray}
% where
%$p_+ = \ptot \frac{1+\beta}{2}$ , $p_- = \ptot \frac{1-\beta}{2}$. 

The likelihood terms $\like^{\mathrm{kin}}_{j}$ describe the kinematic distributions of the $\mpipi^2$, $\beta$, and \ptot\ variables
 for the physics signals and are obtained from Monte Carlo simulations. %~\cite{Ruffini-Thesis}.  
The corresponding distributions for the combinatorial background are extracted from data~\cite{Ruffini-Thesis}
and are included in the likelihood through the $\like^{\mathrm{kin}}_{\mathrm{c}}$ term.
%In particular, the squared-mass distribution of the combinatorial background is parametrized by an exponential function. 
%In the fit, the slope is fixed to the value extracted from an enriched sample of two
%random tracks, containing events passing all requirements of the final selection 
%except for vertex quality replaced by an antiselection criterion  %anti-selection cut  
%$\chi^2 > 40$, which efficiently %strongly 
%rejects track pairs originating from a common vertex.
The likelihood term $\like^{\mathrm{kin}}_{\mathrm{p}}$ describes the kinematic distributions %of $\mpipi^2$, $\beta$, and \ptot\ variables
of the background from partially reconstructed decays of generic $b$ hadrons~\cite{Ruffini-Thesis, Aaltonen:2011jv}.
%The $\mpipi^2$ distribution is modeled by an 
%empirical threshold function,  defined as %$\mpipi^2 \sqrt{1- ( \mpipi^2 / x_0 )^2} $ if $\mpipi^2 < x_0$,
%$x \sqrt{1- ( x / x_0 )^2} $ if $x < x_0$ (where $x=\mpipi^2$),
%convoluted with a Gaussian resolution function,
%where the cutoff $x_0$ is determined from data.
%The $\beta$ and \ptot\ distributions for the physics background 
%are obtained from Monte Carlo simulation. 
%A detailed description of the fit %and its parameters 
%can be found in Refs.~\cite{Ruffini-Thesis, Aaltonen:2011jv}.

%-----------------------------------------------------------------------------------------

To ensure the reliability of the search for small signals in 
the vicinity of larger structures, the shapes of the mass distributions assigned to each signal are
modeled in detail %.  Momentum dependence and non--Gaussian resolution tails are accounted for by 
with the full simulation of the detector.  %~\cite{CDFMC},
%while the 
Effects of soft photon radiation in the final state are simulated by {\sc photos}~\cite{photos}.
The mass resolution model is %accurately 
tuned to the observed shape of the $3.8\times 10^{6}$ \DKpi\ and  $1.7 \times 10^{5}$ \Dpipi\ candidates
in a sample of  $D^{*+}\to  D^0\pi^+$ decays, collected with a similar trigger selection. 
The accuracy of the procedure is checked by comparing the observed mass line-shape 
of $9 \times 10^5$ $\Upsilon(1S) \to \mu^+ \mu^-$ decays
to that predicted by the tuned simulation. %, obtaining discrepancies below the 2\% level.  
A good agreement %($\chi^2/{\rm dof}= 1.15$)  
is obtained when a global scale factor to the mass resolution of $1.017$ is applied to the model. 
Based on this result, we conservatively assign a 2\% systematic
uncertainty to the mass line-shape model.

Particle identification is achieved by means of the energy deposition measurements (\dedx) from the drift chamber.
%Particle identification is achieved by using energy deposition measurements (\dedx) from the drift chamber.
The $D^{*+}\to D^0\pi^+$ sample is also used to calibrate the \dedx\ response %of the drift chamber 
to positively and negatively charged kaons and pions, using the charge 
of the pion from the $D^{*\pm}$ decay to determine the flavour of the neutral $D$ meson. 
%to positively and negatively charged kaons and pions, using the charge 
%of the pion from the $D^{*+}$ decay to identify the $D^0$ decay products. %The \dedx\ 
The response for protons and antiprotons is determined from a sample of $1.4 \times 10^6$ 
$\Lambda\to p \pi^{-}$ decays, where the kinematic properties and the momentum threshold of the trigger allow unambiguous identification 
of the decay products~\cite{Ruffini-Thesis}. 
%%%%
%PID information is summarized by a single observable kaonness, defined as:
%\begin{equation}
%\kappa = \frac{{  d}E/{  d}x - {  d}E/{d}x(\pi)}{{  d}E/{  d}x(K) - {  d}E/{  d}x(\pi)}
%\end{equation}
%where ${  d}E/{  d}x(\pi)$ and ${  d}E/{  d}x(K)$ are the expected ${  d}E/{  d}x$ depositions for those particle assignments.
%%%%%
%%%%
The PID information is summarized by a single observable $\kappa$, defined as follows:

\begin{equation}
\kappa \equiv \frac{{  d}E/{  d}x - {  d}E/{d}x(\pi)}{{  d}E/{  d}x(K) - {  d}E/{  d}x(\pi)},
\end{equation}
in which ${  d}E/{  d}x(\pi)$ and ${  d}E/{  d}x(K)$ are the average expected
specific ionizations given the particle momentum 
for the pion and kaon mass hypothesis, respectively. 
%The statistical separation in the momentum range of interest between kaons and pions 
The statistical separation  between kaons and pions with momentum larger than 2~\pgev\
%The statistical separation power between kaons and pions  with $p_T>2$ \pgev\ 
is about 1.4$\sigma$, while the ionization rates of protons and kaons are quite similar. 
Thus, the separation between $K^{+}\pi^{-}$ or $p\pi^{-}$ final states and their charge-conjugates 
is about 2.0$\sigma$ and 2.8$\sigma$ respectively, while  %~\cite{Ruffini-Thesis}, while 
that between $pK^{-}$ and $\bar{p}K^{+}$ is  %lower, 
about 0.8$\sigma$. %~\cite{Ruffini-Thesis}. 
However, in the last case  
additional discrimination at the 2$\sigma$ level is provided by kinematic differences in 
$(\mpipi^2,\beta)$ distributions~\cite{Aaltonen:2011qt,Ruffini-Thesis}.  %, about 2$\sigma$~\cite{Ruffini-Thesis}.
The PID likelihood term, which is similar for physics signals and backgrounds, 
depends only on $\kappa$ and on its expectation value $\langle \kappa \rangle$ (given a mass hypothesis) for the decay products.
The physics signal model is described by the likelihood term $\mathcal{L}_{j}^{\mathrm{PID}}$, 
where the index $j$ uniquely identifies the final state.
The background model is described 
by the two terms $\mathcal{L}_{\mathrm{p}}^{\mathrm{PID}}$ and $\mathcal{L}_{\mathrm{c}}^{\mathrm{PID}}$, 
respectively, for the physics and combinatorial background, that 
account for all possible pairs that can be formed combining only charged pions and kaons.
%Since background from muons(protons) is order of few per cent,  and since the available \dedx\ resolution 
%to distinguish them from pions(kaons) is limited, they are therefore included within the nominal pion component. 
With the available \dedx\ resolution,
muons are indistinguishable from pions with the available \dedx\ resolution and are therefore included in the pion component.
%In addition the proton component has been also included within the nominal kaon component, because the limited 
%\dedx\ resolution and its small relative fraction, which is of the order of few per cent.  
Similarly, the small proton component in the background is included in the kaon component.
% Thus the physics background model allows for independent, charge-averaged contributions 
% of pions and kaons, whose fractions are determined by the fit;
% while the combinatorial background model, instead, allows for more contributions, since independent fractions of 
%  positively and negatively charged pions and kaons are determined by the fit. 
%
Thus, the combinatorial background model allows for independent positively and negatively charged contributions 
of pions and kaons, whose fractions are determined by the fit, while %instead 
the physics background model, where charge asymmetries are negligible, 
only allows for charge-averaged contributions.

\begin{figure}[!h]
\begin{center}
\begin{overpic}[width=4.2cm,height=4.2cm]{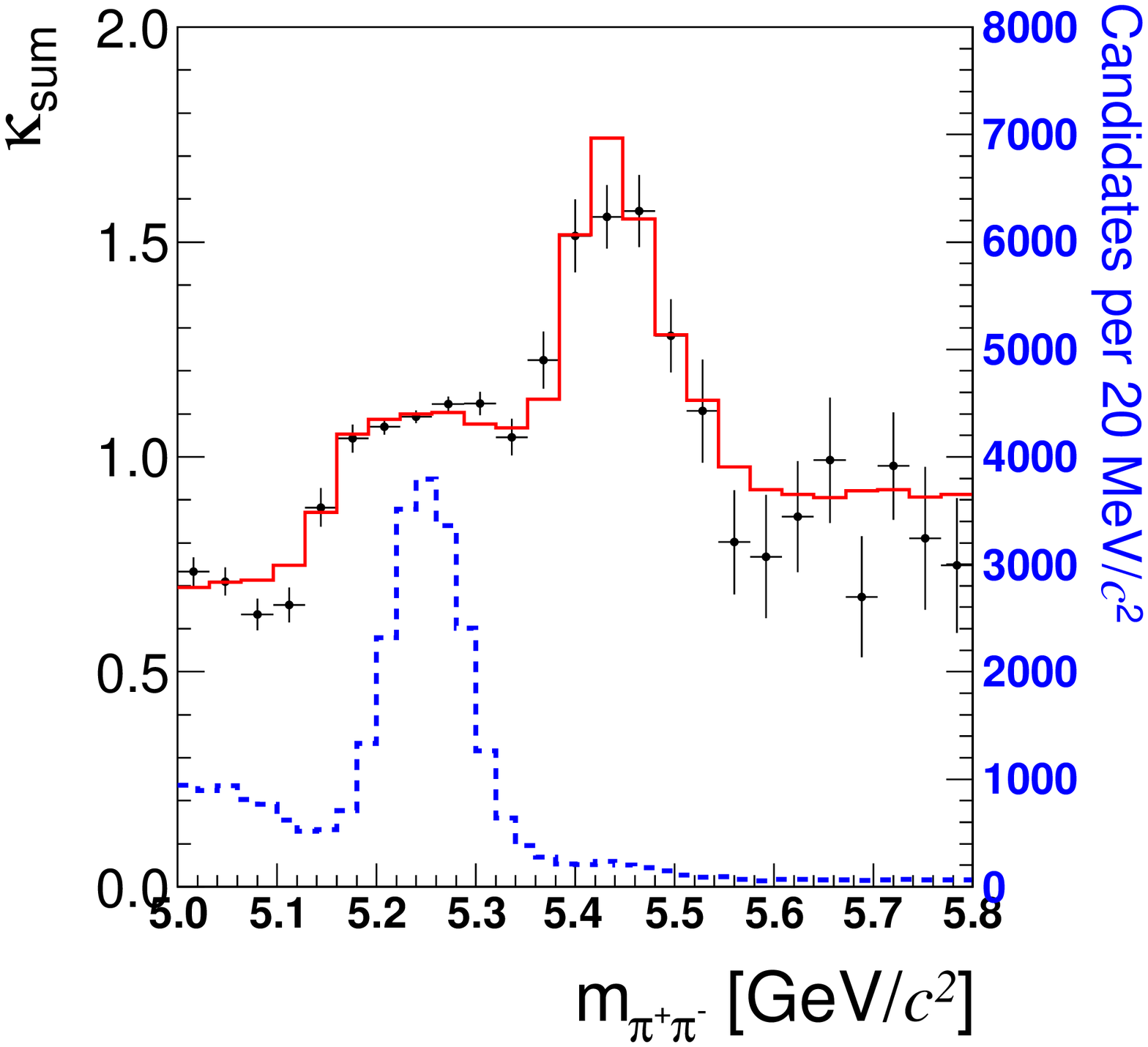}
\put(25,96){(a)}
%\put(44.0,98){\tiny $\it B^0 \to K\pi$}
\multiput(44.0,17.8)(0,1.99){45}{\line(0,1){1}}
%\put(44.0,17.5){\line(0,1){89}} % BdKpi 5.234 --> 5.240
%\put(52.9,90){\tiny $\it B^0_s \to K\pi$}
\multiput(54.0,17.8)(0,1.99){45}{\line(0,1){1}}
%\put(52.9,17.5){\line(0,1){89}}%BsKpi  5.329 --> 5.331
%\put(64.5,36){\tiny $\it \Lambda^0_b \to pK$}
\multiput(64.5,17.8)(0,1.99){45}{\line(0,1){1}}
%\put(64.5,17.5){\line(0,1){89}} %LbpK  5.4378 -->5.437
%\put(69.7,26){\tiny $\it \Lambda^0_b \to p\pi$}
\multiput(69.7,17.8)(0,1.99){45}{\line(0,1){1}}
%\put(69.7,17.5){\line(0,1){89}} %Lbppi 5.491 -->5.491
\end{overpic}   
%\hspace*{0.2cm}
\begin{overpic}[width=4.2cm,height=4.2cm]{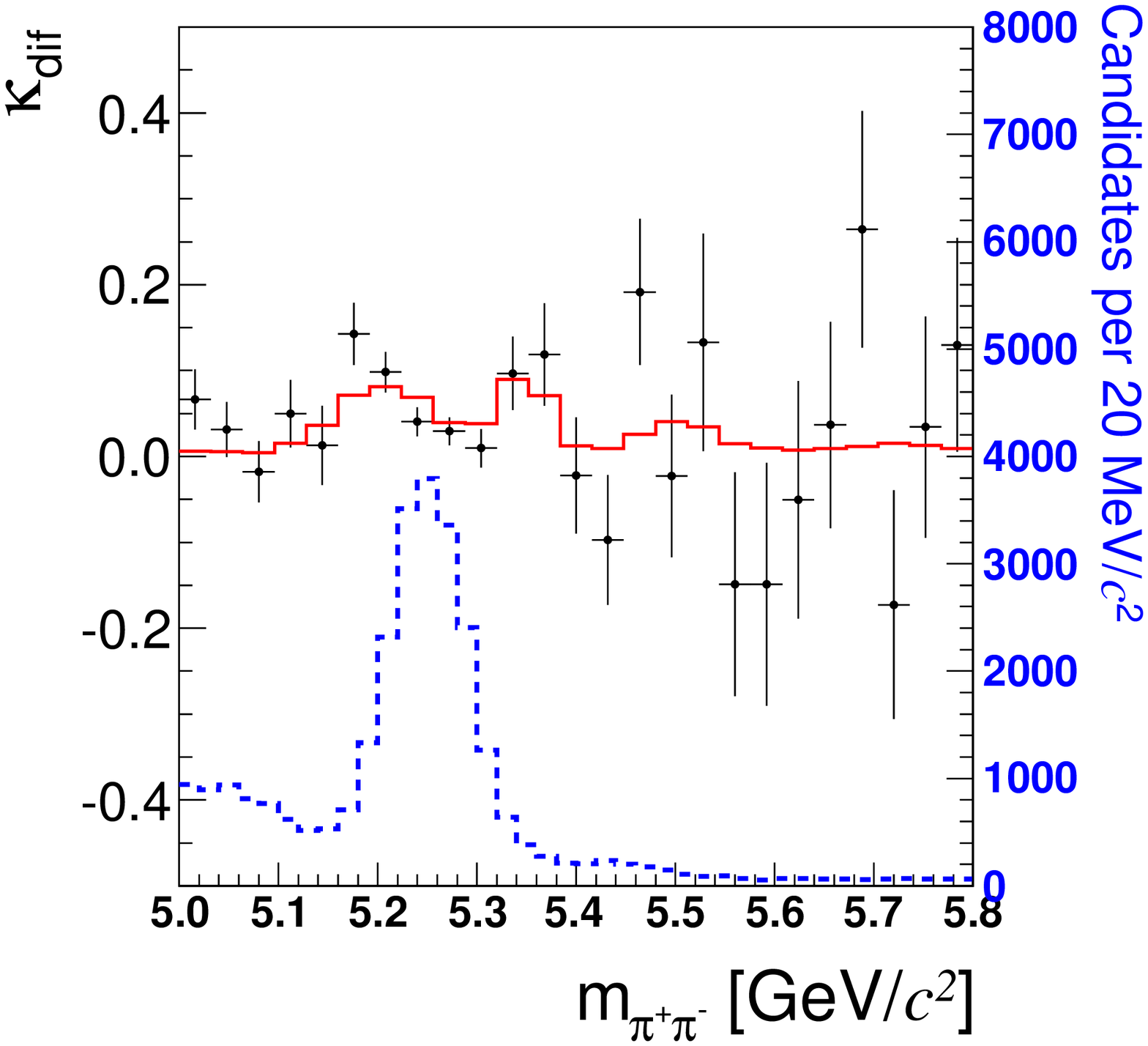}
\put(25,96){(b)}
%\put(44.0,98){\tiny $\it B^0 \to K\pi$}
\multiput(44.0,17.8)(0,1.99){45}{\line(0,1){1}}
%\put(44.0,17.5){\line(0,1){89}} % BdKpi 2.234 --> 5.240
%\put(52.9,90){\tiny $\it B^0_s \to K\pi$}
\multiput(54.0,17.8)(0,1.99){45}{\line(0,1){1}}
%\put(52.9,17.5){\line(0,1){89}}%BsKpi  5.329 --> 5.331
%\put(64.5,36){\tiny $\it \Lambda^0_b \to pK$}
\multiput(64.5,17.8)(0,1.99){45}{\line(0,1){1}}
%\put(64.5,17.5){\line(0,1){89}} %LbpK  5.4378 -->5.437
%\put(69.7,26){\tiny $\it \Lambda^0_b \to p\pi$}
\multiput(69.7,17.8)(0,1.99){45}{\line(0,1){1}}
%\put(69.7,17.5){\line(0,1){89}} %Lbppi 5.491 --> 5.491
\end{overpic}   
\end{center}
\caption{ 
Distribution of the average value of $\kappa_{\rm sum}$  (a) and $\kappa_{\rm dif}$   (b) as a function of  $\mpipi$.
The fit function is overlaid. For reference, the distribution of $\mpipi$ is shown by the dashed lower
histogram. Dashed vertical lines indicate the position, from left to right, of the following signals: \BdKpi, \BsKpi, \LbpK, \Lbppi.  
}
\label{fig:sum_diff_kaonness_plots}
\end{figure}

%
%-------------------------------------------------
%
%   Fit results 
%
%-------------------------------------------------
%
%The results of the fit are in agreement with those obtained 
%from previous measurements on smaller subsamples~\cite{Aaltonen:2008hg, Aaltonen:2011jv}.
To check the goodness of the fit with regard to the PID observables, Fig.~\ref{fig:sum_diff_kaonness_plots} shows the distributions 
of the average value of $\kappa_{\rm sum}=\kappa_{+} + \kappa_{-}$ and $\kappa_{\rm dif}=\kappa_{+} - \kappa_{-}$ as a function 
of $\mpipi$, with fit projections overlaid, where $\kappa_{+}(\kappa_{-})$ is the PID observable for positively(negatively) charged particles.
The $\kappa_{\rm sum}$ distribution is sensitive to the identity of final-state particles, and reveals the presence of 
baryons as a narrow structure, where the mass distribution lacks prominent features.
%appears almost featureless.
Conversely, the $\kappa_{\rm dif}$ distribution is expected to be uniformly zero, %unless
except in the presence of a charge asymmetry 
coupled with a different \dedx\ response of the final particles. It is insensitive to the \LbpK\ signal due to the similarity of 
proton and kaon \dedx\ responses, but it is sensitive to the \CP\
asymmetries of the other decay modes, and indeed it displays a 
deviation %in correspondence of 
corresponding to each of the other three decay modes object of this study.
The signal yields from the likelihood fit of Equation~(\ref{like_fit}) are reported in Table \ref{tab:ACPs} together with the physical asymmetries, \acpbtof, derived as follows:

\begin{equation}
\ACPRAWdef = 
\frac{N_{\btof} - c_{f}N_{\abtoaf}}{N_{\btof} + c_{f}N_{\abtoaf}},
\end{equation}
where $c_{f} = \varepsilon(f)/\varepsilon(\bar{f})$ is the ratio
between the efficiencies for triggering and reconstructing 
the final states $f$ and $\bar{f}$. %with respect to the state $\bar{f}$. 
The $c_{f}$ factors correct for detector-induced charge asymmetries and are extracted from control samples in data. 
Simulation is used only to account for differences between the kinematic distributions of $H_b \to h^+h'^{-}$ decays and control signals. 

%--------------------------------------------------------------------------------------------------
%
%   ACP table     
%
%--------------------------------------------------------------------------------------------------
%\begin{table*}
\begin{table}[!h]
 \caption[Results]{%Direct \CP\ asymmetries. 
CP-asymmetry results. % Direct \CP-violating asymmetries results. 
 The first quoted uncertainty is statistical; the second is systematic. 
 $\mathcal{N}$ is the number of events determined by the fit for each decay mode. }
 {\footnotesize
 \begin{tabular}{lccc}
 \hline \hline
% Decay &  $\mathcal{N}_{\btof}$ & $\mathcal{N}_{\abtoaf}$ & \acp(\btof)(\%)   \\
%  \hline
% \BdKpi             & 5313 $\pm$ 109      & 6348 $\pm$ 117         & $-$8.3 $\pm$ 1.3 $\pm$ 0.4           \\
%  \BsKpi            & 560 $\pm$ 51          & 354   $\pm$ 46           &  $+$22 $\pm$ 7 $\pm$ 2             \\
% \Lbppi             & 242   $\pm$ 24         & 206 $\pm$ 23             &  $+$6 $\pm$ 7 $\pm$ 3      \\ 
%  \LbpK             & 271   $\pm$ 30        & 324 $\pm$ 31             &  $-$10 $\pm$ 8 $\pm$ 4      \\
%%%%
Decay &  $\mathcal{N}_{\btof}$ & $\mathcal{N}_{\abtoaf}$ & \acp(\btof)   \\
\hline
\BdKpi             & 5313 $\pm$ 109      & 6348 $\pm$ 117         & $-0.083$ $\pm$ 0.013 $\pm$ 0.004           \\
\BsKpi            & 560 $\pm$ 51          & 354   $\pm$ 46           &  $+0.22$ $\pm$ 0.07 $\pm$ 0.02             \\
\Lbppi             & 242   $\pm$ 24         & 206 $\pm$ 23             &  $+0.06$ $\pm$ 0.07 $\pm$ 0.03      \\ 
\LbpK             & 271   $\pm$ 30        & 324 $\pm$ 31             &  $-0.10$ $\pm$ 0.08 $\pm$ 0.04      \\
%%%%
\hline \hline
 \end{tabular}
 }
 \label{tab:ACPs}
\end{table}
% \end{table*}
%------------------------------------------------------------------------------------------------------------------------
The corrections for $f=K^+\pi^-$ are extracted from a sample of 
3$\times$10$^{7}$ $D^0\rightarrow K^-\pi^+$ decays collected without requiring the
$D^{*+}\to  D^0\pi^+$ decay chain~\cite{Aaltonen:2011se}. 
By imposing the same offline selection to the $D^0$ decays, we obtain $K^{\mp}\pi^{\pm}$ 
final states in a similar kinematic regime to that of the $H_b$ signals.
%to the signals. %(see Fig.~\ref{fig:kaonness_pdf}). 
We assume that $K^+\pi^-$ and $K^-\pi^+$ final states from charm decays 
are produced in equal numbers %at the Tevatron 
because their production is dominated by the strong interaction and, compared to the detector effects
to be corrected, the possible \CP--violating asymmetry in $D^0\to K^-\pi^+$ 
decays is tiny $(<10^{-3})$ as predicted by the SM \cite{Bianco:2003vb}. 
%and confirmed by current experimental determinations~\cite{babar_and_belle_acp_d0}. 
We also check that possible asymmetries in $D^0$ meson yields induced 
by \CP\ violation in $B \to DX$ decays are small and can be neglected~\cite{Aaltonen:2011se}.
Therefore, any asymmetry between observed numbers of reconstructed 
$K^-\pi^+$ and $K^+\pi^-$ charm decays is ascribed to detector-induced 
effects and used to extract the desired correction factor. 
The ratio $N_{\aDKpi}/N_{\DKpi}$ is measured by performing a simultaneous fit to 
the invariant $K^-\pi^+$ and $K^+\pi^-$ mass distributions~\cite{Aaltonen:2011se}.
%We find a $c_{K^{+}\pi^{-}}= 1/c_{K^{-}\pi^{+}}=1.011\pm0.001$,  
%consistent with its previous estimate~\cite{Aaltonen:2011qt}. 
%%%%
We find a significant asymmetry $c_{K^{+}\pi^{-}}= 1/c_{K^{-}\pi^{+}}=1.011\pm0.001$,  
consistent with expectation %due to
%its previous estimate~\cite{Aaltonen:2011qt}, 
based on charge asymmetries of the interaction probability with detector material~\cite{Acosta:2004ts}. %giagu
%due to the different probability of strong interaction of charged partticles in 
%in the tracker material between positively and negatively charged kaons.} % and pions.}
%consistent with its previous estimate
%\textcolor{red}{from data~\cite{Aaltonen:2011qt}  and with estimate done using the full detector simulation, 
%that accounts for the different probability of strong interaction in the tracker material between positively and negatively charged kaons and pions}.
%ter we describe the measurement of the detector-induced charge asymmetry between posi tively and negatively charged kaons and pions,
% due to their different probability of strong interaction in the tracker material. 
%%%%
We also add a systematic uncertainty 
that allows for a possible nonvanishing \CP\ violation, using the 
available experimental knowledge \acp(\DKpi) = ($0.1\pm0.7$)\%~\cite{PDG}. 
For the \Lbppi\ asymmetry, the factor 
$c_{p\pi^{-}}$ is extracted from data using a similar strategy, where 
a simultaneous binned $\chi^2$ fit to the $\Lambda \to p \pi^-$ and $\bar{\Lambda} \to \bar{p} \pi^+$
mass distributions is performed to estimate observed yields~\cite{Ruffini-Thesis}.
We average the obtained value with the same estimate based on simulation, taking half the difference as a systematic uncertainty. 
The final value is $c_{p\pi^{-}}=  1.03 \pm 0.02$~\cite{Ruffini-Thesis}.
In the measurement of \CP\ violation in \LbpK\ decays, 
instrumental charge asymmetries
induced from both kaon and proton interactions are relevant.
The $c_{pK^{-}}$ factor is determined by the product  %by multiplying $c_{p\pi^{-}} \cdot c_{K^{-}\pi^{+}}$ 
$c_{p\pi^{-}} \cdot c_{K^{-}\pi^{+}}$
based on the assumption that the efficiency $\varepsilon(f)$ %for two particles
factorizes as the product of the single-particle efficiencies.

%-----------------------------------------------------------
 %   Systematics   
%-----------------------------------------------------------

The dominant systematic uncertainties on \acp(\Lbppi) and \acp (\LbpK) are due to the uncertainty on the 
model of the momentum distributions of %momentum %($\beta$,\ptot) templates of 
the combinatorial background and the lack of the knowledge on the  $\Lambda^0_b$ spin-alignment.
A polarized initial state would affect 
the distributions of the momentum-related variables used in the fit.
A systematic uncertainty is assessed 
by repeating the fit accounting for a nonvanishing polarization, by taking the difference with the the central fit done in the hypothesis of no polarization.
The dominant contribution to the systematic uncertainty on \acp (\BdKpi) %is due to the \dedx\ model. This 
originates from the statistical uncertainty in the parameters used to model %analytically  
the correlated \dedx\ response of  the two decay products~\cite{Ruffini-Thesis}. 
%This uncertainty is evaluated by repeating the likelihood fit of composition 500 times with various sets of \dedx\ parameters, randomly extracted 
%as Gaussian-distributed variables, correlated as in data, and centered
%at the central value of the parametrization.
%The resulting systematic uncertainty on each observable is obtained as the three-standard deviation range of the 
%distribution of that observable over the ensemble of fits. 
In the case of \acp (\BsKpi), the systematic uncertainty mainly originates 
from three sources of similar importance: %having the same size: %.  It comes %for \acp (\BsKpi) 
the uncertainty on the background and signal kinematic templates, 
the uncertainty on the \dedx\ modeling discussed above, 
and the uncertainty on trigger efficiencies.

 %-----------------------------------------------------
%   Final comments   
 %-----------------------------------------------------
Table \ref{tab:ACPs} reports the final results, that are consistent with and supersede the previous CDF results~\cite{Aaltonen:2011qt}.
The asymmetries of the \LbpK\  and  \Lbppi\ modes are now more precisely determined by a factor of 2.3 and 2.0, respectively.  % and are unique. 
%The resolution on the asymmetries of the \LbpK\  and  \Lbppi\ modes are improved by more than a factor of two.
%The asymmetries of the \LbpK\  and  \Lbppi\ modes are unique on the field and themeasurements and their resolution 
%Table \ref{tab:ACPs} reports the final results, that are consistent with and supersede the previous CDF results~\cite{Aaltonen:2011qt}. 
%The asymmetries of the \LbpK\  and  \Lbppi\ modes are now more precisely determined by a factor of nearly three.  % and are unique.
These are unique measurements. Both results are consistent with zero, excluding a
large \CP\ asymmetry in these decay modes, 
which was predicted by calculations~\cite{Lu:2009cm}  that yielded negative asymmetries for \Lbppi\ of approximately 30\%,
albeit with large uncertainties. 
The same calculation also predicts a  vanishing asymmetry for the
\LbpK, implying a predicted difference $\acp(\Lbppi)-\acp(\LbpK)
\approx -0.26 $ between the two modes, 
to be compared to the measurement $0.16\pm 0.12$. 
The uncertainty on the theory prediction is not known; it is a
difference between two numbers with large uncertainties, 
but they are likely to be at least partially correlated. Evaluating
this correlation would allow a more useful comparison with the experimental value.

We confirm the observation of \acp(\BdKpi) with a  significance larger than $5\sigma$.
The measured value is consistent with the latest results from asymmetric $e^+e^-$ colliders~\cite{acp_bfactories} and  
LHCb~\cite{Aaij:2013iua}. 
We also find a nonzero \acp(\BsKpi) with a significance of 3.0$\sigma$, in good agreement with % This result confirms 
the recent LHCb measurement $\acp(\BsKpi) = +0.27 \pm 0.04\stat \pm 0.01\syst$~\cite{Aaij:2013iua}, thus providing 
confirmation of their first observation of \CP\ violation in the \Bs-meson system.  % dynamics.
The simultaneous measurement of \CP\ asymmetries in the $\Bd$ and $\Bs$ meson decays to $K^{\pm}\pi^{\mp}$ final states
allows a quantitative test of the SM-prediction $\acp(\BsKpi)= +0.29 \pm 0.06$ \footnote{We use the relation $\acp(\BsKpi) = -\acp(\BdKpi) 
\frac{\mathcal{B}(\BdKpi)}{\mathcal{B}(\BsKpi)}  \frac{\tau(\Bs)}{\tau(\Bd)}$ from Ref.~\cite{Gronau:2000md}.},
consistent with our measurement at the 10\% level. %of $\acp(\BsKpi)$ 
%is consistent at the 10\% level with this prediction, \textcolor{red}{$\acp(\BsKpi) = +0.29 \pm 0.06$}, 
This is obtained using the world average of the decay rates and lifetimes~\cite{PDG}
of the two decay modes, assuming SM origin of the \CP\ violation in these channels and U-spin symmetry.

 %-----------------------------------------------------
 %    Conclusions   
 %-----------------------------------------------------

%In summary, we report the final CDF measurements of the \CP\ asymmetries of charmless neutral $b$-hadrons %$b-$meson and $b-$baryon 
%decays into pairs of charged hadrons, using the complete Run II data sample. 
%All the results contribute significantly to the world averages. % and are unique for the baryon modes.
%These are unique measurements for the baryon decays.

In summary, we report the final CDF measurements of the \CP\ asymmetries of charmless neutral $b$-hadrons 
%$b-$meson and $b-$baryon 
decays into pairs of charged hadrons, using the complete Run II data sample.
We confirm the observation of \acp(\BdKpi) with a  significance larger than $5\sigma$, and we find 
 a nonzero \acp(\BsKpi) with a significance of 3.0$\sigma$. Results on $b-$baryon decays %contribute significantly to the world averages. % and are unique for the baryon modes.
$\acp(\Lbppi) = +0.06 \pm 0.07\stat \pm 0.03\syst$ and $\acp(\LbpK) = -0.10 \pm 0.08\stat \pm 0.04\syst$, 
are unique measurements and are compatible with no asymmetry. 
 %for the baryon decays.

\begin{acknowledgments}
We thank the Fermilab staff and the technical staffs of the
participating institutions for their vital contributions. This work
was supported by the U.S. Department of Energy and National Science
Foundation; the Italian Istituto Nazionale di Fisica Nucleare; the
Ministry of Education, Culture, Sports, Science and Technology of
Japan; the Natural Sciences and Engineering Research Council of
Canada; the National Science Council of the Republic of China; the
Swiss National Science Foundation; the A.P. Sloan Foundation; the
Bundesministerium f\"ur Bildung und Forschung, Germany; the Korean
World Class University Program, the National Research Foundation of
Korea; the Science and Technology Facilities Council and the Royal
Society, United Kingdom; the Russian Foundation for Basic Research;
the Ministerio de Ciencia e Innovaci\'{o}n, and Programa
Consolider-Ingenio 2010, Spain; the Slovak R\&D Agency; the Academy
of Finland; the Australian Research Council (ARC); and the EU community
Marie Curie Fellowship Contract No. 302103.

\end{acknowledgments}

 % % Create the reference section using BibTeX:
 % %\bibliography{basename of .bib file}
 % 
 % %Uncomment the following two lines to use BibTeX
%  \bibliography{bibliografiav4}

\begin{thebibliography}{37}
\bibitem[{CDF()}]{CDF-coordinates}
\emph{\bibinfo{title}{{\em CDF II uses a cylindrical coordinate system in which
 $\phi$ is the azimuthal angle, $r$ is the radius from the nominal beam line,
 and $z$ points in the proton beam direction, with the origin at the center of
 the detector. The transverse plane is the plane perpendicular to the $z$
 axis.}}}
\end{thebibliography}
%  \end{document}
 %

%\begin{thebibliography}{99}

%\end{thebibliography}

\end{document}